\documentclass[11pt,twoside]{article}
\usepackage{./asp2014}

\aspSuppressVolSlug
\resetcounters

\begin{document}

\title{OHANA, the AMBER/VLTI Snapshot Survey}
\author{Th.~Rivinius,$^{1}$ W.J.~de Wit,$^{1}$ Z.~Demers,$^{2,3}$ 
         A.~Quirrenbach,$^{3}$  and the VLTI Science Operations Team
\affil{$^{1}$ESO - European Organisation for Astronomical Research in the
  Southern Hemisphere, Chile; email: {\tt triviniu@eso.org}}
\affil{ $^{2}$Department of Physics and Astronomy, The University of Western
  Ontario, London, Canada}
\affil{$^{3}$Landessternwarte K\"onigstuhl, D-69117, Heidelberg, Germany}}

% This section is for ADS Processing.  There must be one line per author.
\paperauthor{Th.~Rivinius}{triviniu@eso.org}{}{ESO - European Organisation for Astronomical Research in the Southern Hemisphere}{}{Vitacura}{}{}{Chile}
\paperauthor{W.J.~de Wit}{wdwit@eso.org}{}{ESO - European Organisation for Astronomical Research in the Southern Hemisphere}{}{Vitacura}{}{}{Chile}
\paperauthor{Z.~Demers}{zdemers@uwo.ca}{}{Department of Physics and Astronomy, The University of Western Ontario}{}{London}{}{}{Canada}
\paperauthor{A.~Quirrenbach}{A.Quirrenbach@lsw.uni-heidelberg.de}{}{Landessternwarte K\"onigstuhl}{}{Heidelberg}{}{}{Germany}

\begin{abstract}
We report on the OHANA interferometric snapshot survey, carried out by the
VLTI group at the Paranal observatory. It makes use of observing time not
useful for any other scheduled scientific or technical tasks in the sense of a
backup programme, to characterize the mass-loss for early-type stars.  The
survey employs the combination of AMBER's high spectral and spatial
resolution. The spatially unresolved central object provides a reference frame
for the fringe properties observed in the light of the continuum.
\end{abstract}

\section{Introduction}
The {\bf O}bservatory survey at {\bf H}igh {\bf AN}gular resolution of {\bf
  A}ctive OB stars (OHANA) combines high spectral with high spatial resolution
across the Br$\gamma$ and He{\sc i}$\lambda$2.056 lines to characterize the
dynamics of winds and disks. It was carried out by the VLTI group at the
Paranal observatory with the three-beam combining instrument AMBER
\citep{2007A&A...464....1P}. The survey was designed to make use of the
observing time not requested by other programs, usually due to bad weather or
unsuitable local sidereal time slots.

\section{Observations and Data Reduction}
The initial survey targets, observed in ESO period P93, consisted of twelve
bright Be stars, thirteen O and B type supergiants, and one interacting binary
(see {Table~1}). Almost 300 observations were obtained. Due to the unforeseen
availability of time P94 during a few weeks, the program was revived for this
limited time and a few targets. However, mainly supergiants to complement the
initial selection and more appropriate to the new range in right-ascension,
were added, and only one additional Be star, Achernar.

By design, namely targeting quantities relative to the adjacent continuum, no
calibrators were observed. However, in some nights calibrators, taken for
technical purpose or other programs using the same setup, are available. These
have been added to the database, too.

Basic data reduction was performed with amdlib, v3.0.6
\citep{2007A&A...464...29T,2009A&A...502..705C}, and then processed further
with idl. The raw data have become public immediately, the final reduction of
the P93 Br$\gamma$ observations is completed and will be made
public. Reduction of the P94 and He{\sc i}$\lambda$2.056 data is pending.

\begin{table}
\begin{center}
\caption{Observed Be stars, their spectral types, and data obtained. For P93
  observations,the number of observations on the {\bf s}mall, {\bf
    i}ntermediate, and {\bf l}arge telescope configurations (s--i--l) are
  given for each spectral line. In P94 only Br$\gamma$ was
  observed}
\label{Rivinius_OHANA_tab1}
\begin{tabular}{llcc}
Target             & Sp.~type   & Br$\gamma$ & He{\sc i}$\lambda$2.056\\
                   &            &  s--i--l   & s--i--l \\
\hline\\[-2ex]
\multicolumn{4}{c}{\bf P93}\smallskip \\
$\mu$\,Cen         & B2\,Vnpe & 2--5--3 & 0--1--0 \\
$\chi$\,Oph        & B2\,Vne  & 0--0--1 & 0--0--0 \\
$\zeta$\,Tau       & B2\,IVe-sh  & 2--1--0 & 1--0--0\\
$\delta$\,Cen      & B2\,IVne & 3--5--2 & 1--1--1\\
$\epsilon$\,Cap    & B3\,Ve-sh& 1--5--4 & 0--2--0\\
$\beta^1$\,Mon\,A  & B3\,Ve   & 6--8--0 & 2--1--0\\
$\beta^1$\,Mon\,B  & B3\,ne   & 2--1--0 & 0--0--0\\
$\beta^1$\,Mon\,C  & B3\,e    & 2--1--0 & 0--0--0\\
P\,Car             & B4\,Vne  & 6--5--2 & 2--3--1\\
$\beta$\,Psc       & B6\,Ve   & 1--4--4 & 0--2--0\\
$\eta$\,Tau        & B7\,IIIe & 0--0--0 & 1--0--0\\
Electra            & B8\,IIIe & 0--0--0 & 1--0--0\smallskip\\
\multicolumn{4}{c}{\bf P94}\smallskip \\
$\chi$\,Oph        & B2\,Vne  & 4 \\
$\epsilon$\,Cap    & B3\,Ve-sh& 7\\
$\alpha$\,Eri      & B4\,Vne  & 8 \\
$\beta$\,Psc       & B6\,Ve   & 5\\

\hline\\[-2ex]
\end{tabular}\smallskip
\end{center}
\end{table}

\section{Data Description and First Impressions}
\begin{figure}[t]
\begin{center}
\parbox{0.5\textwidth}{\centerline{$\mu$\,Cen (B2\,Vnpe)}}%
\parbox{0.5\textwidth}{\centerline{P\,Car (B4\,Vne)}}%

\includegraphics[width=0.5\textwidth]{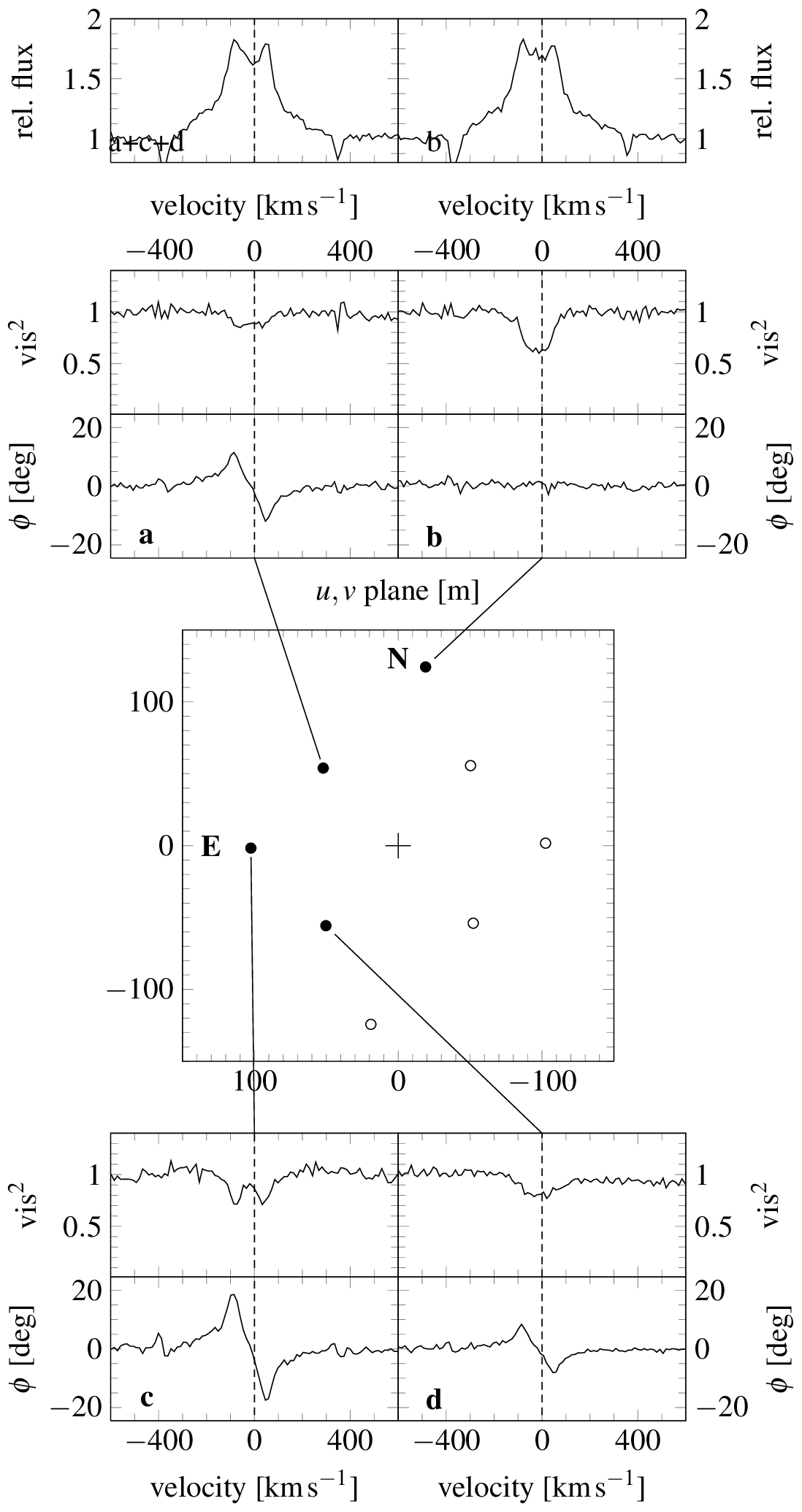}%
\includegraphics[width=0.5\textwidth]{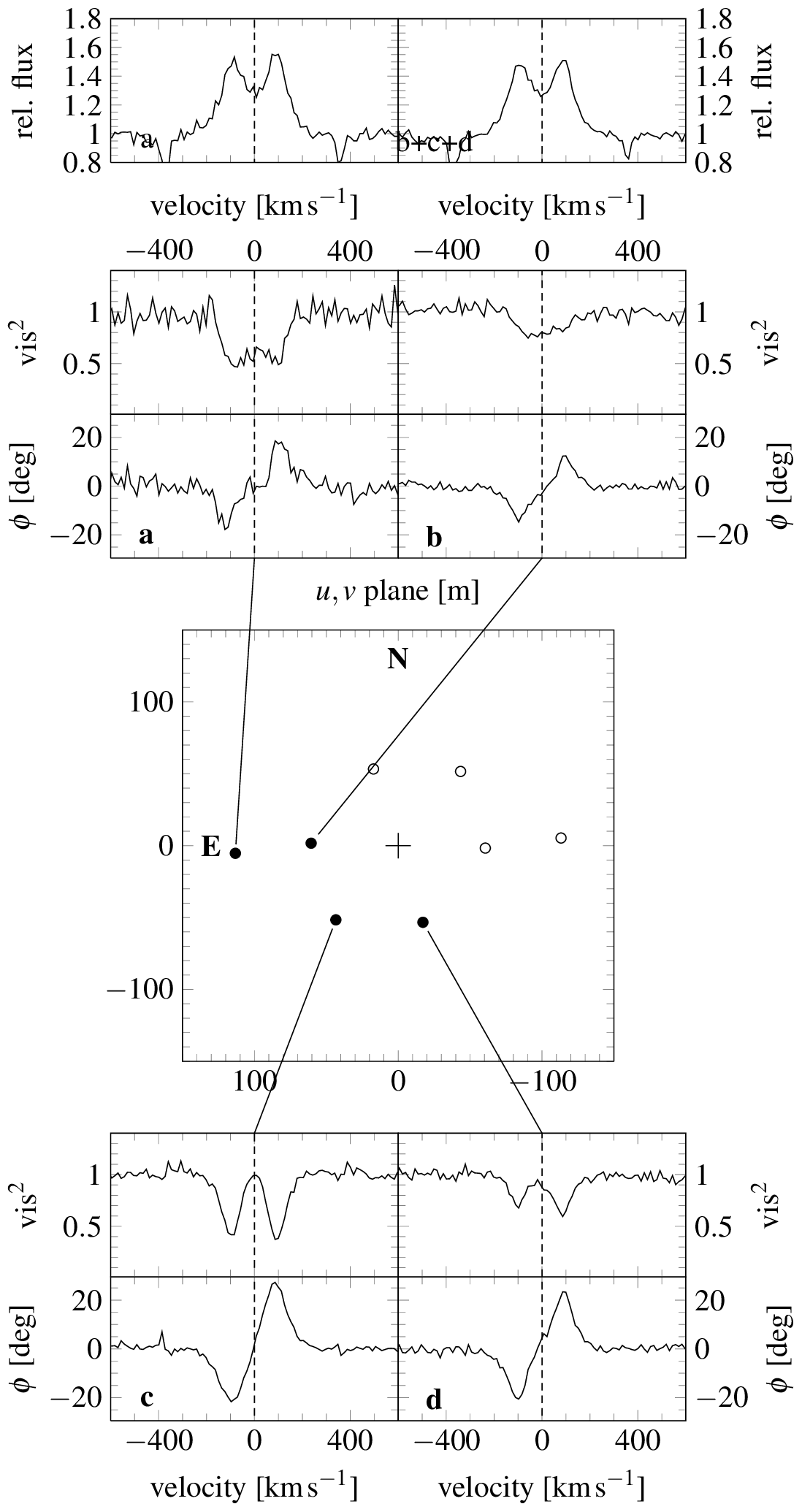}%
\caption{Example of OHANA data for the Be stars $\mu$\,Cen and P\,Car.}
\label{Rivinius_OHANA_fig1}
\end{center}
\end{figure}

\begin{figure}[t]
\begin{center}
\parbox{0.5\textwidth}{\centerline{$\delta$\,Cen (B2\,IVne)}}%
\parbox{0.5\textwidth}{\centerline{$\chi$\,Oph (B2\,Vne)}}%

\includegraphics[width=0.5\textwidth]{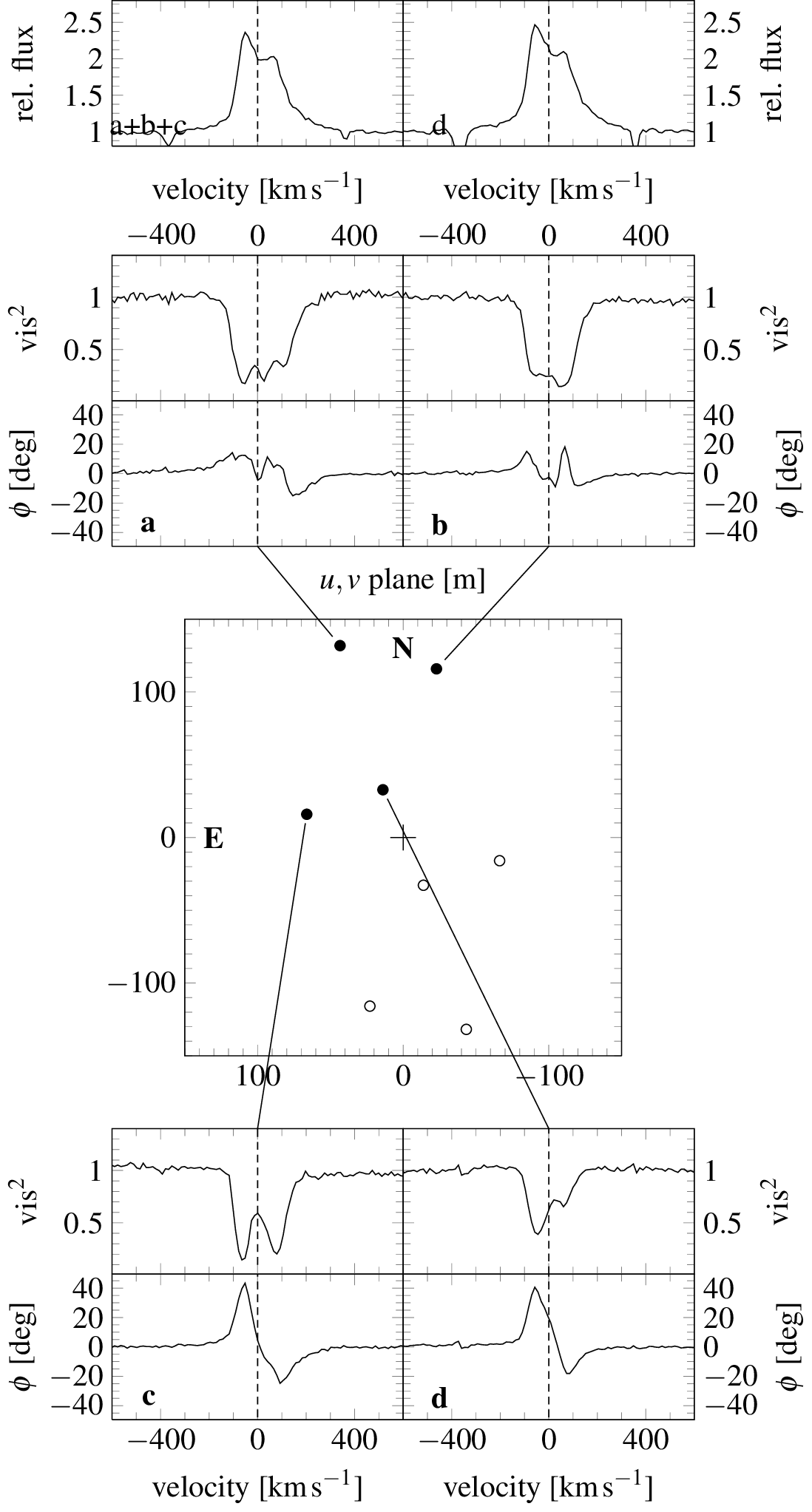}%
\includegraphics[width=0.5\textwidth]{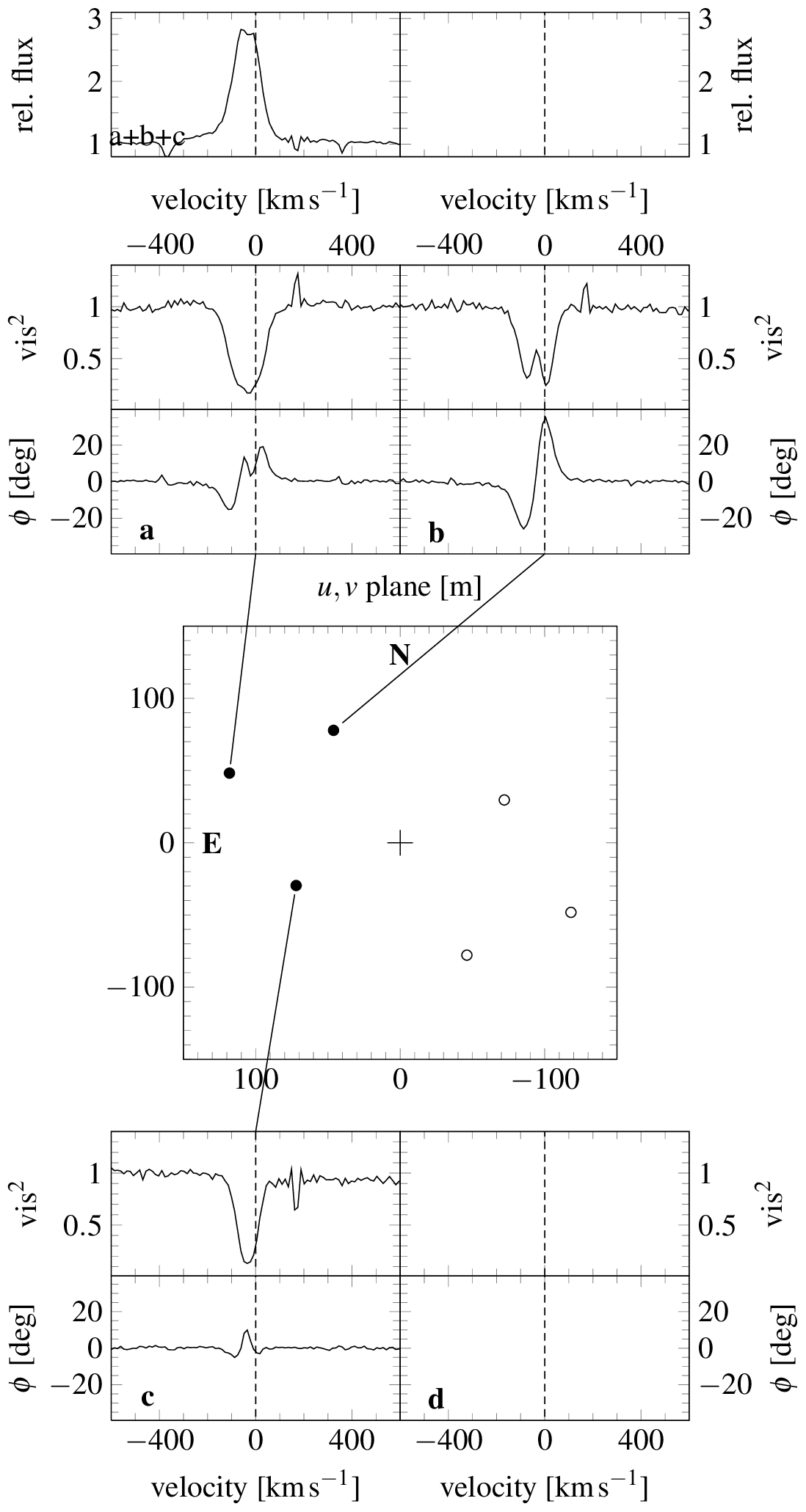}%
\caption{Example of OHANA data for the Be stars $\delta$\,Cen and $\chi$\,Oph.}
\label{Rivinius_OHANA_fig2}
\end{center}
\end{figure}

Due to the snapshot/backup/filler nature of the program, the data quality is
inhomogeneous. Typical values for a good data set are an uncertainty of the
visibility (normalized to unity) of about $\pm0.05$, and of the phase $\pm
2^\circ$, at a SNR of the combined spectrum of above 100.
Selected data sets of the target stars are shown in
Figs.~\ref{Rivinius_OHANA_fig1} and \ref{Rivinius_OHANA_fig2}. For each of the
four targets, four baselines are shown, taken from two observations. The
uppermost panels for each target show the flux spectra, then subpanels a-d show
visibility and phase (upper and lower resp.\ profiles), while the centered
panel show the ($u,v$) plane covered by the four baselines shown.

Visual inspection of the Be star observations shows them to be {compatible
  with the canonical picture}, namely a cicumstellar decretion disk. The
targets span all inclinations (equatorial to pole-on) and spectral
subtypes. For some of the brighter stars the disk is already well resolved in
the intermediate configuration (typical baseline lengths 30--70m), and overly
resolved in the large configuration (typical baseline lengths 80--130m) Data
for {$\beta^1$\,Mon and $\mu$\,Cen are shown in
  Fig.~\ref{Rivinius_OHANA_fig1}}.  $\mu$\,Cen shows a broad shallow ramp-type
wing in the line, which is reflected in the phase. This may be the signature
of freshly ejected material closer to the star than the bulk of the disk.

\section{Conclusions}
The OHANA survey provided interferometric data of the circumstellar
environments of Be stars and OBA supergiants. The raw data is publicly
available, the reduced data will become so as soon as the final reduction has
passed quality control tests. The reduced data will be made available at:\\
{\tt http://activebstars.iag.usp.br/index.php/34-ohana}.

\acknowledgements Based on observations made with ESO Telescopes at the La Silla Paranal Observatory under programme ID 093.D-0298

%\bibliography{editor}  % For BibTex
\bibliography{OHANA}

\end{document}